# Applications and Use of Microemulsions

Kai Lun LEE, CID:00664757

*Department of Chemical Engineering and Chemical Technology,
Imperial College London, November 2010*

**Abstract**

During the past five decades since the discovery of microemulsions by Jack H. Shulman, there have been huge progresses made in applying microemulsion systems in a plethora of research and industrial processes. While it is beyond the scope of this paper to give a complete review of all significant developments and applications, it will attempt to highlight several recent developments in applications that might interest readers for whom this paper serves as an introduction to microemulsions. In that note, the relevance of this paper and the truncated scientific background on microemulsions are first discussed.

**INTRODUCTION**

Microemulsions, or µ-emulsions, are isotropic mixtures of oil, water and surfactant; usually with a co-surfactant and the oil being a mixture of different hydrocarbons and olefins[1]. In contrast to ordinary emulsions which are kinetically stable but are thermodynamically unstable and will phase separate[2], microemulsions are thermodynamically stable and therefore do not require high inputs of energy or shear conditions for their formation. They are also clear, as compared to emulsions which are cloudy. Despite being named microemulsions, the droplet sizes of the dispersed phase in microemulsions are generally in the magnitude of ~10nm. Apart from being more commercially viable due to its lower energy requirements, microemulsions are also attracting the interest of researchers due to their potential as drug delivery vehicles, in other food and pharmaceutical applications, and in the petrochemical industry.

**RELEVANCE**

The significance and potential that researchers attach to microemulsions are in no small part due to their unique properties that are low interfacial tension, high thermodynamic stability, high interfacial area, and the ability to dissolve immiscible liquids.

These have allowed researchers to discover, over the past few decades, many practical uses for microemulsions that are employed in many industrial processes and consumer products, either as new processes and products or as more efficient processes and better or cheaper products. Examples described in this paper include drugs, cosmetics, electronics, and petrol. In short, microemulsions have and can continue to enable us to make better consumer products. There is particularly strong interest in the potential of microemulsions as novel drug delivery systems, as they are believed to hold the key to making better orally delivered drugs that can be absorbed better, and only at selected lengths along the gastrointestinal tract. With pharmaceutical revenue reaching

---

[1] Chandra, A(1992). *Microemulsions: An Overview* Retrieved from http://www.pharmainfo.net/reviews/microemulsions-overview

[2] K, Shinoda. B,Lindman (1987). *Organised surfactant systems: microemulsions*, Langmuir 3 135–149

497 billion USD in 2007, the monetary incentive alone is considerable.

**BACKGROUND**

The concept of microemulsions was introduced by Professor, Jack H. Shulman at Columbia University in 1959[3]. The definition of a microemulsion have varied with time and location, but the more commonly accepted view is that of a "system of water, oil and amphiphile which is a single optically isotropic and thermodynamically stable liquid solution"[4] and that in a microemulsion the surfactant is located at a certain boundary between the oil and aqueous phases, giving the microemulsion a definite microstructure.

The abovementioned surfactant molecules, in most cases, comprise of a polar head which make up a small fraction of the molecular volume and a non-polar tail. These two regions allow them to interact with the polar aqueous phase and the non-polar oil phase. Surfactant molecules associate into different forms, including spherical micelle, rod micelle laminar phase and hexagonal phase, to minimise the gibbs free energy of the system and to "optimise solvation requirements".[5]

The formation and stability of microemulsions can be described by interface science, chemical solubility theories, or by thermodynamic explanations; the last one of which shows that a net release of free energy is obtained when favourable entropic contributions from the mixing of small droplets in the continuous phase and the diffusion of surfactant in the inter-facial layer are larger than the unfavourable contribution of the reduction of surface tension[5], resulting in a thermodynamically stable dispersion.

Microemulsions are usually characterised by ternary phase diagrams, which three edges are the components of a microemulsion, namely, oil, water and surfactant. Any co-surfactant used are usually grouped together with the surfactant at a fixed ratio and treated as a pseudo-component.

The below shows a hypothetical ternary phase diagram of a microemulsion; illustrating that 2 phase systems forms at very high surfactant concentrations.

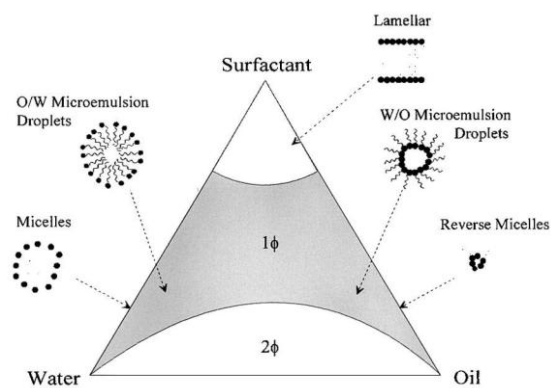

Fig 1. Tenary phase diagram of microemulsion (courtesy of Advanced Drug Delivery Reviews)

Co-surfactants are usually used in conjunction with surfactants due to most single chain surfactants being incapable of reducing the interface tension of oil and water to form a microemulsion. The most common co-surfactants are medium chain alcohols, which reduce the tension and increase the fluidity of the oil-water interface, thereby increasing the entropy of the system[6]. These medium chain alcohols also increase the motility of the surfactants' non-polar tail region, allowing greater penetration by oil molecules and therefore stabilising the system and facilitating the formation of a microemulsion.

In spite of their ease of formation, it is important to accurately characterise microemulsions, especially for industrial processes. Characterisation of microemulsions includes both macroscopic measurements and microenvironment methods. The former include viscosity measurements which indicates the presence (or absence) of certain surfactant forms, conductivity measurements which can determine the continuous and dispersed phases, as well as dielectric measurements which gives insight to the structure and dynamics of the particular microemulsion[7]. On the other hand, microenvironment studies can involve pulsed field NMR and scattering methods such as light scattering, neutron scattering and X-ray scattering.

---

[3] J.H. Schulman, W. Stoeckenius, L.M. Prince, *Mechanism of formation and structure of micro emulsions by electron microscopy*, J. Phys. Chem. 63 (1959) 1677–1680

[4] I. Danielsson, B. Lindman, *The definition of a microemulsion, Colloids and Surfaces 3* (1981) 391–392

[5] M.J. Lawrence, G.D. Rees, *Microemulsions-based media as novel drug delivery systems*, Advanced Drug Delivery Reviews 45 (2000) 89–121

[6] M.J. Lawrence, *Surfactant systems: microemulsions and vesicles as vehicles for drug delivery,* Eur. J. Drug Metab. Pharmacokinet. 3 (1994) 257–269.

**RECENT DEVELOPMENTS**

<u>Water in Oil Drug Delivery Microemulsions</u>

The rationale behind exploring water in oil (w/o) microemulsions is to protect water soluble drug molecules, in particular proteins and peptides from metabolism and to overcome physical barriers. These are particularly attractive as such drug molecules are heat-sensitive and these emulsions do not require high temperatures to form.

More interestingly, w/o microemulsions with encapsulated Active Pharmaceutical Ingredients (APIs) are found to convert into oil in water (o/w) microemulsions upon the addition of particular aqueous fluids[8], which results in the release of the API. This allows w/o microemulsions to be designed to selectively release APIs at required locations along the gastrointestinal tract. Typical microemulsions used for this purpose incorporate fatty-acid esters as the oil phase and isopropanol as the co-surfactant. Studies carried out using these microemulsions successfully showed that the absorption of vasopressin in the intestines of rats was tripled when compared to using aqueous solutions; and also resulted in much higher bioavailability of insulin in the colon of dogs when a colon-release microemulsion was designed and used[9].

Microemulsions employed in the form of oral-delivered drugs can come with a self-emulsifying ability. These are called Self-microemulsifying drug delivery systems (SMEDDSs). Such SMEDDS particularly interest researchers because of their ability to deliver hydrophobic drugs. With almost 40% of new drug compounds hydrophobic[10], the potential commercial viability of SMEDDS is great.

In spite of the promise of increased dissolution and bioavailability of APIs in oral forms, w/o microemulsions and SMEDDSs have not been commercially or industrially exploited. This is in part due to the lack of knowledge as to how drug molecules are distributed between the oil and aqueous phase, metabolism of the oil phase, and most importantly, how the absorption in human bodies (as compared to in animals) of oral dosage drugs are affected if the oil is non-digestible[11]. This depends on the combination of the drug and the oil phase (usually a lipid), and can increase, inhibit or not affect absorption. The difficulty in understanding these processes arises from the *in vivo* behaviour of drugs due to several physical and physiological factors beyond the scope of this report[11].

That being said, further research and insight on the relationship among drug structures, microemulsion compositions and absorption can resolve current drug development issues and exploit the advantages of microemulsions in drug delivery systems.

---

[7] Y. Feldman, N. Kozlovich, I. Nir, N. Garti, *Dielectric spectroscopy of microemulsions, Colloids Surfaces*: Physicochem. Eng. Aspects 128 (1997) 47–61.

[8] A. J. Owen, S. H. Yiv, A. Sarkahian, *Convertible microemulsion formation* 29 October 1992

[9] W. A. Ritschel. *Microemulsions for improved peptide absorption from the gastrointestinal tract.* Meth. Find. Exp. Clinic. Pharmacol. 13:205-220 (1993)

[10] M. J. Patel, S. S. Patel, N. M. Patel *A Self Microemulsifying Drug Delivery System* International Journal of Pharmaceutical Sciences Review and Research Volume 4 Issue 3 (Sept 2010)

[11] P. P. Constantinides *Lipid Microemulsions for Improving Drug Dissolution and Oral Absorption:* Pharmaceutical Research Vol 12 (1995)

## Water in Carbon Dioxide Microemulsions

Supercritical Carbon Dioxide is a "useful replacement for organic solvents to minimise waste and volatile organic carbon emissions" as it is "non-flammable, non-toxic and the least expensive solvent after water."[12] The apparent advantages of carbon dioxide over organic solvents have been tempered by difficulties in dispersing water in carbon dioxide to form a stable water-in-$CO_2$ microemulsion. Recent efforts have, in particular, been geared towards making $CO_2$ accessible to proteins and similar hydrophilic chemicals through water in $CO_2$ microemulsions.

Theoretically, a dispersion of a hydrophilic or lipophilic phase in $CO_2$ could be stabilised by surfactants with "$CO_2$-philic" tails and hydrophilic or lipophilic heads respectively. The difficulty however, was finding such a surfactant that could disperse water in $CO_2$. Attempts to locate such a surfactant involved experimenting with 150 different substances in the decade preceding[13] the 1996 discovery that commercially available ammonium carboxylate perflouropolyether (PFPE) with an average molecular weight of 740 are suitable surfactants for water in $CO_2$ microemulsions at pressures below 300bar[13]. Johnston and team (1996) noted that the use of PFPE created a thermodynamically stable aqueous dispersed phase in continuous $CO_2$ that more importantly, did not alter the three dimensional conformation of proteins which are important to their functions; having demonstrated the results through the use of Fourier Transform Infrared Spectroscopy, ultraviolet absorbance, fluorescence and electron paramagnetic resonance experiments.

Other surfactants with similar attributes such as having tail groups with weak van der Waal's forces and low solubility in water, an ionic head group with a tendency to leave $CO_2$, and most importantly, carbon fluoride branching which bends the interface of water can possibly act as suitable surfactants for water in $CO_2$ microemulsions.

## Synthesis of Nanosized Bismuth Oxyhalide Particles and other nanoparticles using microemulsions

Bismuth Oxyhalides are crucial to a number of applications including BiOCl in cosmetics and in cracking of butane and BiOI as "components for color filters in transparent nanocomposite materials"[14]. Like other nanosized particles, their shape, size and size distribution are crucial to their applications in "catalysis, electronics, miniaturization, and ceramics[14]".

Reverse microemulsions are often used in the preparation of nanoparticles to avoid broad particle size distributions that often result from precipitation. A reverse microemulsion differs from an emulsion in that the tail groups of the surfactants orientate outwards into the continuous phase and their head groups towards the dispersed phase.

A relatively new method of using microemulsions to synthesis the nanoparticles is through a multi-microemulsion route. Such a route involves sets of microemulsions with the same water, oil and surfactant used and similar ratios of the 3 components, but with each of the reactants for the nanoparticles dissolved in the aqueous phase of different microemulsions. Intermicellar exchange of reactants present in the reverse micelles subsequently occurs and the nanoparticles form. (See illustration on next page.) Studies show that "the multi-microemulsion route

---

[12] K.P.Johnston et al *Water-in-Carbon Dioxide Microemulsions: An Environment for Hydrophiles Including Proteins* Science, New Series, Vol. 271, No. 5249 (Feb. 2, 1996), pp 624-626

[13] K. A. Consani and R. D. Smith, *J. Supercritical Fluids 3*, 51 (1990).

[14] J. Henle et al. *Nanosized BiOX (X) Cl, Br, I) Particles Synthesized in Reverse Microemulsions* Chem. Mater. 2007, pp366-373

yields finer particles and narrow size distributions.[15]"

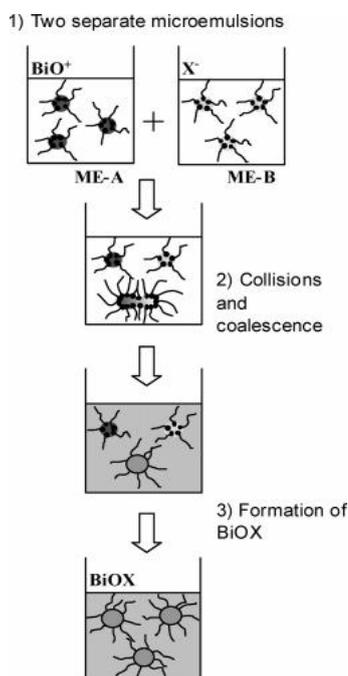

Fig 2. Multi-microemulsions route of synthesising nanoparticles. Courtesy of Henle(2007)[14]

By varying the concentrations of the salt solutions in the aqueous phase and the water to surfactant ratio, Henle and team (2007) managed to characterize and formulate the relationship between those conditions and the particle size in the range of 3 to 22nm; allowing them to manipulate characteristics of the nanoparticles such as "the band gap, absorption edge, and color of BiOI nanoparticles".

Further on, processing and preparations of other nano-sized substances also employed the use of microemulsions such as for nanocrystalline hydroxyapatite[16] and for single crystal BaMoO4 nanofibers[17].

The ability to narrow the size distribution, create finer nanoparticles, and control their size using microemulsions also found an application in the study of reaction kinetics in catalytic processes. Some reaction kinetics and mechanism are dependent on the catalyst and/or its size and well-defined catalysts in the nanometer range deposited on supports are very often indispensible to study them.

It has been found that water-in-oil microemulsions are particularly advantageous over other methods of preparing nanoparticle catalysts on supports, such as electron beam lithography, colloidal lithography and spin-coating, as they "can be formed at atmospheric pressure and at room temperature and that large sample volumes relatively easily can be obtained.[18]" This process involves adding support to the microemulsion suspension and subsequently destabilising the microemulsion by the addition of a destabilizer such as tetrahydrofurane which removes the surfactant.

Enhanced Oil Recovery using Microemulsions

The use of microemulsions is of high interest in many aspects of crude oil exploitation, but none more so that in enhanced oil recovery. In cases where the pressure exerted by gushing sea water on the oil phase is not able to overcome capillary forces sufficiently, microemulsions are the key to extracting more than just a minor portion of crude oil. Properly balanced microemulsions are able to do so by drastically reducing the interfacial tension to the magnitude of 0.001 mN m$^{-1}$ [19]. This is also known as chemical flooding.

Many difficulties are encountered in creating a suitable microemulsion with temperature gradients required large, many ionic surfactants precipitating when contacted with brine, and most non-ionic surfactants

---

[15] Wang, J. A et al. Today 2001, 68, pp21.
[16] L. Hong et al *Processing of nanocrystalline hydroxyapatite particles via reverse microemulsions*, Journal of Materials Science, v 43, n 1, p 384-389, January 2008
[17] Z. Li et al *Synthesis of single crystal BaMoO4 nanofibers in CTAB reverse microemulsions* Materials Letters, v 59, n 1, p 64-68, January 2005

[18] H.H.Ingelsten et al *Deposition of Platinum Nanoparticles, Synthesized inWater-in-Oil Microemulsions,* Langmuir 2002, 18, pp1811-1818
[19] L. Lake*, Enhanced Oil Recovery,* Prentice-Hall Inc, 1989.

unsuitable[20]. Other issues include adsorption of microemulsion components on rocks, and varying salinities and temperatures of the sea. Over the past three decades, however, there has been significant progress made on the recovery of residual oil in particular chemical flooding with microemulsions and many patents issued, mostly concerned with "the chemistry of surfactant-based processes including the use of chemical solutions to decrease the surface tension between oil and the flooding medium, screening of surfactants for oil recovery efficiency, chemical surfactant designs and formulation to mobilize residual oil and other factors in the chemistry of surfactant-based chemical flooding processes.[21]"

Recent reviews have focused on the application of suitable microemulsion systems in specifically conditions of the North Sea leading to many patents filed by various companies involved in oil recovery there; while researchers are also excited by the possible use of bacteria and biosurfactants that are able to create microemulsions. These substitutes are, in addition to being able to multiply within capillaries, capable of physically modifying solid surfaces and lowering viscosity of the oil phase[22], and are viewed as having great potential of lowering recovery costs.

**CONCLUSION**

Microemulsions, a relatively recent discovery in 1959, have found applications in a wide variety of chemical and industrial processes. Their wide use in both research and industry is in part due to their ease of formation and stability.

This paper has highlighted several recent developments in the use or studies on the use of microemulsions both for industrial and research purposes. It is regretted that it is beyond the scope and length of this paper to highlight all outstanding recent developments involving the use of microemulsions and is therefore limited only to the discussions on the application of microemulsions in novel drug delivery systems, use of $CO_2$ as a solvent, in enhanced oil recovery and in the synthesis of nanoparticles. In the first development discussed, microemulsions enable the selective release Active Pharmaceutical Ingredients along specified lengths in the gastrointestinal tract by inverting from a water-in-oil emulsion to an oil-in-water emulsion. In the second, the discovery of a suitable surfactant allows $CO_2$ to form a microemulsion in water thereby possibly eliminating the need to use organic solvents which may be toxic, flammable and expensive. In the third, microemulsions enable a much larger fraction of highly valuable residual oil trapped under sea beds to be recovered by reducing interfacial tensions. Research on biosurfactants, which additionally may multiply themselves and lower viscosity of the oil phase, is also ongoing. In the last development discussed, nanoparticles with wide ranges of applications in industry, research and consumer products are able to be made finer, with a narrower size distribution and better controlled properties by using microemulsions in their production.

The field of microemulsion applications is far from being fully established and researchers face many difficulties yet are intrigued by their immense potential in many areas due to their extremely low interfacial tension, thermodynamic stability, large interfacial area and ability to dissolve immiscible liquids. With continued research and studies, microemulsions could possibly play a much bigger role in our lives by affecting the ways industries work and research is conducted.

---

[20] J. Sjiiblom, R. Lindbergh, S.Friberg *Microemulsions - phase equilibria characterization, structures, applications and chemical reactions* Advances In Colloid and Interface Science 1996 pp125-287

[21] B.Y.Jamaloei *Insight to the chemistry of surfactant-based enhanced oil recovery* January 2009

[22] B.K.Paul, S.P.Moulik *Uses and applications of microemulsions* Current Science 80 April 2001